\documentstyle[12pt]{article}
\begin{document}
 \title{Split-Quaternions and the Dirac Equation}
\author{Francesco Antonuccio\footnote{\texttt{f\_antonuccio@yahoo.co.uk}} \\
 London, United Kingdom}
 \renewcommand{\today}{July, 2014}
 \maketitle
 \abstract 
We show that Dirac 4-spinors admit an entirely equivalent formulation in terms of 2-spinors defined over the split-quaternions. In this formalism, a Lorentz transformation is represented as a $2 \times 2$ unitary matrix over the split-quaternions. The corresponding Dirac equation is then derived in terms of these 2-spinors. In this framework the $SO(3,2; {\bf R})$ symmetry of the Lorentz invariant scalar $\overline{\psi}\psi$ is manifest.   

\section {Introduction}
It is a well known fact that there are no finite unitary representations of the Lorentz group over the complex numbers. In this paper we give an example of a  simple finite unitary representation of the Lorentz group over the split-quaternions. The split-quaternion algebra is a non-commutative extension of the complex numbers, and admits the operation of conjugation. The conjugate of a split-quaternion reduces to the usual definition of complex conjugation when restricting to the subalgebra of complex numbers.

The split-quaternion algebra has a long history (see \cite{sqn} for the original work by James Cockle), and 
the suggestion that (split-)quaternionic analysis and representation theory has a role to play in a modern understanding of four dimensional physics has been discussed in a series of mathematical papers by Frenkel and Libine \cite{Frenkel}, \cite{FrenkelII}, \cite{Libine}.  More recently, the author noted a connection between split-quaternions and mappings from spinors to spacetime \cite{AntonuccioSpinorPaperIII}.

In this paper, we focus our attention on Dirac 4-spinors and the accompanying Dirac equation. In particular, we show that the usual Dirac 4-spinor $\psi$ with complex components may be replaced by a 2-spinor $\Psi$ with split-quaternionic components. Moreover, in this new formalism, we prove that the Dirac equation takes the following  form:
\begin{equation}
\label{eq:sqDiracEquation}
\left( H^{\mu} \partial_{\mu} + m \right) \cdot \Psi = 0,
\end{equation}
where 
\begin{equation}
\label{eq:HMatrices}
\begin{array}{cccc}
H^0 = \left(
\begin{array}{cc}
{\rm i} & 0 \\
0 & {\rm i}
\end{array}
\right) &
H^1 = \left(
\begin{array}{cc}
{\rm j} & 0 \\
0 & {\rm j}
\end{array}
\right) &
H^2 = \left(
\begin{array}{cc}
{\rm k} & 0 \\
0 & {\rm -k}
\end{array}
\right) &
H^3 = \left(
\begin{array}{cc}
0 & {\rm k} \\
{\rm k} & 0
\end{array}
\right) .
\end{array}
\end{equation}

The elements ${\rm i}$, ${\rm j}$ and ${\rm k}$ appearing in the matrices (\ref{eq:HMatrices})  are the anti-commuting generators of the split-quaternion algebra, and will be defined in Section \ref{sec:SQ}. This number system is particularly useful since it admits the operation of conjugation. In particular, a split-quaternionic number multiplied by its conjugate is always real-valued, but unlike the quaternions, it can take on negative values.

We will also be interested in matrices defined over this non-commutative algebra, so we will need to make precise the rules of matrix multiplication, and define the conjugate transpose (i.e. Hermitian conjugate) of a split-quaternionic matrix. This occurs in Section \ref{sec:SQM}. With these definitions in place, we can study the Lie algebra of unitary matrices over the split-quaternions, and deduce that the transformation properties of a Dirac 4-spinor with complex components has an equivalent formulation in terms of a 2-component spinor defined over the split-quaternions. This is proved in Section \ref{sec:SQSpinors}. In this new formalism, the Lorentz group in four dimensional spacetime is a six dimensional subgroup of the $2 \times 2$ unitary matrices over the split-quaternions. The Lie algebra of this unitary group turns out to be isomorphic to the ten dimensional Lie algebra of $SO(3,2; {\bf R})$. 

In Section \ref{sec:SQDiracEqn} we strengthen this equivalence between Dirac 4-spinors and split-quaternionic 2-spinors by showing that Dirac's equation admits a simple split-quaternionic formulation.

We summarise our results and discuss applications in Section \ref{sec:Conclusions}. 

\section{The Split-Quaternions ${\bf P}$}
\label{sec:SQ}
\subsection{Definitions}
\label{sec:defnsq}
Only a basic understanding of the algebraic properties of the split-quaternions is required for this article. The reader is referred to \cite{sq2},\cite{sq3},\cite{sq4} and \cite{sq5} for some references. An element $p$ of the split-quaternion algebra ${\bf P}$ is any number that has the form
\begin{equation}
\label{eq:sqdefn}
p = a_1 + {\rm i} a_2 + {\rm j} a_3 + {\rm k} a_4
\end{equation}
where the $a_i$ are real numbers, and ${\rm i}, {\rm j}$ and ${\rm k}$ are anti-commuting generators satisfying the following algebra:
\begin{equation}
\label{eq:sqalgebra}
\begin{array}{ccc}
{\rm i}^2 = -1 &
{\rm j}^2 =+1 &
{\rm k}^2 = +1 \\
{\rm i}{\rm j} = +{\rm k} & 
{\rm j}{\rm k} = -{\rm i} &
{\rm k}{\rm i} = +{\rm j} \\
{\rm i}{\rm j} = -{\rm j}{\rm i} &
{\rm j}{\rm k} = -{\rm k}{\rm j} &
{\rm k}{\rm i} = -{\rm i}{\rm k} 
\end{array}
\end{equation}
Using the relation ${\rm k} = {\rm i}{\rm j}$ above, we can rewrite the definition (\ref{eq:sqdefn}) as
\begin{equation}
\label{eq:sqdefncomplex}
p = z_1 + z_2 {\rm j}
\end{equation}
where $z_1$ and $z_2$ are complex numbers, and 
\begin{equation}
\label{eq:jsquared}
{\rm j}^2 = +1.
\end{equation}
Also, the relation ${\rm i}{\rm j} = -{\rm j}{\rm i}$ implies
\begin{equation}
\label{eq:sqbarrelation}
z{\rm j} = {\rm j}\overline{z}
\end{equation}
for any complex number $z$. In fact, we can define the split-quaternions by the relations (\ref{eq:sqdefncomplex}),  (\ref{eq:jsquared}) and (\ref{eq:sqbarrelation}), where addition and multiplication are defined in the obvious way. It is now straightforward to prove that this alternative definition of the split-quaternion algebra is entirely equivalent to the standard definition specified by relations (\ref{eq:sqdefn}) and (\ref{eq:sqalgebra}).

In any case, it follows that the split-quaternions form an associative and distributive algebra under the rules of addition and multiplication. 

Now, given any split-quaternion $p= a_1 + {\rm i} a_2 + {\rm j} a_3 + {\rm k} a_4$, we define the conjugate $\overline{p}$ by writing 
\begin{equation}
\label{eq:sqconjugation}
\overline{p} = a_1 - {\rm i} a_2 - {\rm j} a_3 - {\rm k} a_4.
\end{equation}
With this definition, we can prove the following identity:
\begin{equation}
\label{eq:quadraticform}
\overline{p}p=p\overline{p}=(a_1)^2+(a_2)^2-(a_3)^2-(a_4)^2.
\end{equation}
Now if $p_1$ and $p_2$ are any two split-quaternions, then 
\begin{equation}
\label{eq:bar1}
\overline{p_1+p_2} = \overline{p_1}+\overline{p_2}  
\end{equation}
and
\begin{equation}
\label{eq:bar2}
\overline{p_1 \cdot p_2} = \overline{p_2} \cdot \overline{ p_1}.
\end{equation}
We also define the `modulus-squared' of a split-quaternion $p$, written $|p|^2$, as follows:
\begin{equation}
\label{eq:modulus2}
|p|^2=\overline{p}p.
\end{equation}
Hence the modulus-squared $|p|^2$ of a split-quaternion $p$ is always real-valued. Note, however, that it can take on negative values, and so differs in character to the corresponding definition for quaternions.

Combining definition (\ref{eq:modulus2}) with identity (\ref{eq:bar2}) above, we deduce the following:
\begin{equation}
\label{eq:modproduct}
|p_1 \cdot p_2|^2 = |p_1|^2 |p_2|^2.
\end{equation}   

As a side remark, if we use the alternative formalism by writing $p= z_1+z_2{\rm j}$ where $z_1$ and $z_2$ are complex numbers, then the above relations can be translated as follows:
\begin{equation}
\overline{p} = \overline{z}_1 - z_2{\rm j},
\end{equation}
and
\begin{equation}
|p|^2 = |z_1|^2-|z_2|^2.
\end{equation}

We are now ready to define the inverse of a split-quaternion. Namely, if $p$ is a split-quaternion satisfying $|p|^2\neq 0$, then 
\begin{equation}
\label{eq:inversesq}
p^{-1} = \frac{\overline{p}}{|p|^2}
\end{equation}
is a well defined and unique inverse for $p$. However, there are an infinite number of non-zero split-quaternions satisfying $|p|^2=0$, and so the split-quaternions form a non-division algebra.

\subsection{The Unit Split-Quaternions and SL$(2,{\bf R})$}
\label{sec:unitsq}
The set of all split-quaternions $p \in {\bf P}$ satisfying 
\begin{equation}
|p|^2 = 1
\end{equation}
forms a group with respect to multiplication. Closure under multiplication and inversion follows from identities (\ref{eq:modproduct}) and (\ref{eq:inversesq}) respectively. By virtue of the identification 
\begin{equation}
\label{eq:SQvscomplexmatrix}
p=z_1+z_2{\rm j} \leftrightarrow
\left(
\begin{array}{cc}
z_1  & z_2 \\
\overline{z}_2 & \overline{z}_1
\end{array}
\right).
\end{equation}
between a split-quaternion $p=z_1+z_2 {\rm j}$ and a $2 \times 2$ complex matrix, and noting that  $|p|^2=|z_1|^2-|z_2|^2$ is just the determinant of the matrix, it follows that the group of unit split-quaternions is isomorphic to the non-compact complex group SU$(1,1;{\bf C})$, which is isomorphic to the special linear group SL$(2,{\bf R})$. 

\section{Matrices over the Split-Quaternions}
\label{sec:SQM}
\subsection{Basic Definitions}
The basic properties of split-quaternionic matrices have been considered before and the reader should consider \cite{sqmatrices1},\cite{sqmatrices2},\cite{sqmatrices3} for a more detailed algebraic treatment. 
Since the split-quaternions form a non-commutative algebra, the multiplication of matrices defined over the split-quaternions needs to be made precise. If $A$ and $B$ are two $n \times n$ matrices with split-quaternionic
entries $A_{ij},B_{ij} \in {\bf P}$, then the product $A \cdot B$ is another $n \times n$ matrix whose entries $(A \cdot B)_{ij} \in {\bf P}$ are defined in the obvious way:
\begin{equation}
\label{eq:sqmproduct}
(A \cdot B)_{ij} = \sum_{k=1}^{n}A_{ik}B_{kj}.
\end{equation}
The conjugate transpose (or Hermitian conjugate) of a split-quaternionic matrix $A$, written $A^{\dagger}$, is defined by simply conjugating each of the entries in accordance with definition (\ref{eq:sqconjugation}), and then transposing the resulting matrix. Formally, 
\begin{equation}
\label{eq:sqmHconjugate}
(A^{\dagger})_{ij} = \overline{A_{ji}}.
\end{equation}
It then follows from identities (\ref{eq:bar1}) and  (\ref{eq:bar2}), and the above definitions (\ref{eq:sqmproduct}) and (\ref{eq:sqmHconjugate}), that for any two $n \times n$ split-quaternionic matrices $A$ and $B$, the following identities hold:
\begin{equation}
(A + B)^{\dagger} = A^{\dagger} + B^{\dagger}
\end{equation}
and
\begin{equation}
\label{eq:sqmdagger}
(A \cdot B)^{\dagger} = B^{\dagger} \cdot A^{\dagger}.
\end{equation}
A simple consequence of the above is the following:
\begin{equation}
\label{eq:sqmndagger}
(A^m)^{\dagger} = (A^{\dagger})^m,
\end{equation}
where $A$ is any $n \times n$ split-quaternionic matrix and $m$ is any positive integer. Keeping in mind that the exponential of an $n \times n$  split-quaternionic matrix $A$, written $e^A$, is defined by the series expansion
\begin{equation}
e^A = {\bf 1} + A + \frac{1}{2!}A^2+\frac{1}{3!}A^3 + \cdots,
\end{equation}
(where ${\bf 1}$ is the $n \times n$ identity matrix), we can make use of (\ref{eq:sqmndagger}) to derive the following important relation:
\begin{equation}
\label{eq:sqmexpdagger}
(e^A)^{\dagger} = e^{A^{\dagger}}.
\end{equation}
This last fact will be useful when studying the Lie algebra of unitary groups generated by split-quaternionic matrices.

As a final remark, we point out that if $A$ is any given $n \times n$ split-quaternionic matrix, then 
\begin{equation}
\label{eq:sqminverse}
e^A \cdot e^{-A} = e^{-A} \cdot e^{A} = {\bf 1}.
\end{equation}
One way to prove this is to first identify a given split-quaternion $p=z_1+z_2{\rm j}$ with a  $2 \times 2$ complex matrix as specified by (\ref{eq:SQvscomplexmatrix}). Using this correspondence, the $n \times n$ split-quaternionic matrix $A$ appearing in (\ref{eq:sqminverse}) may be replaced by a $(2n) \times (2n)$ complex matrix. A simple application of the Baker-Campbell-Hausdorff formula then yields the required result.

\subsection{The Unitary Group U$(2,{\bf P})$}
\label{sec:u2p}
For a given positive integer $n$, we define U$(n,{\bf P})$  to be the set of all $n \times n$ matrices over the split-quaternions that are unitary with respect to the Hermitian conjugate operator ${}^{\dagger}$ defined by (\ref{eq:sqmHconjugate}). Formally, $U$ $\in$ U$(n,{\bf P})$ if and only if 
\begin{equation}
\label{eq:sqmunitarity}
U^{\dagger} \cdot U = U \cdot U^{\dagger} = {\bf 1}.
\end{equation}
It is easy to check that U$(n,{\bf P})$ forms a group under the operation of matrix multiplication. For the simple case $n=1$, we end up with the group of unit split-quaternions, which was discussed in Section \ref{sec:unitsq}. In particular, we made the observation there that
\begin{equation}
\mbox{U$(1,{\bf P})$ $\cong$ SU$(1,1; {\bf C})$ $\cong$ SL$(2,{\bf R})$ }.
\end{equation}
For the purposes of this article, we will focus our attention on the group U$(2,{\bf P})$ of $2 \times 2$ unitary matrices, since this will contain as a subgroup a spinor representation of the Lorentz group.

Towards this end, let us consider those elements $U$ $\in$ U$(2,{\bf P})$ that may be written in the form 
\begin{equation}
U = e^{As}
\end{equation}
where $A$ is a $2 \times 2$ split-quaternionic matrix and $s$ is an auxiliary real parameter.  By virtue of identity (\ref{eq:sqmexpdagger}), the unitarity condition (\ref{eq:sqmunitarity}) implies
\begin{equation}
\label{eq:sqmunitarity2}
e^{A^{\dagger} s} e^{As} = {\bf 1}.
\end{equation}
Differentiating both sides of equation (\ref{eq:sqmunitarity2}) with respect to the real parameter $s$, and then setting this parameter to zero, we arrive at the following anti-Hermitian constraint on $A$:
\begin{equation}
\label{eq:lieconstraint}
A^{\dagger} = -A.
\end{equation}
It is now straightforward to show that if $A$ is any $2 \times 2$ matrix over the split-quaternions satisfying constraint (\ref{eq:lieconstraint}), then it can always be decomposed as follows:
\begin{equation}
\label{eq:u2pdecomp}
A = \sum_{i=1}^{3}\alpha_i J_i + \sum_{i=1}^{3}\beta_i K_i + \sum_{i=0}^{3}\eta_i L_i
\end{equation}
where the $\alpha_i,\beta_i$ and $\eta_i$ appearing above are real parameters, and the $J_i,K_i$ and $L_i$ are ten matrices specified below:
\begin{equation}
\label{eq:Jmatrices}
\begin{array}{ccc}
J_1 = \frac{1}{2}\left(
\begin{array}{cc}
0 & 1 \\
-1 & 0
\end{array}
\right)
&
J_2=  \frac{1}{2}\left(
\begin{array}{cc}
0 & {\rm i} \\
{\rm i} & 0
\end{array}
\right)
&
J_3=  \frac{1}{2}\left(
\begin{array}{cc}
{\rm- i} & 0 \\
0 & {\rm i}
\end{array}
\right)
\end{array}
\end{equation}
\begin{equation}
\label{eq:Kmatrices}
\begin{array}{ccc}
K_1 =  \frac{1}{2}\left(
\begin{array}{cc}
{\rm -k} & 0 \\
0 & {\rm -k}
\end{array}
\right)
&
K_2=  \frac{1}{2}\left(
\begin{array}{cc}
{\rm j} & 0 \\
0 & {\rm -j}
\end{array}
\right)
&
K_3= \frac{1}{2} \left(
\begin{array}{cc}
0 & {\rm j} \\
{\rm j} & 0
\end{array}
\right)
\end{array}
\end{equation}
\begin{equation}
\label{eq:Lmatrices}
\begin{array}{cccc}
L_0 =  \frac{1}{2}\left(
\begin{array}{cc}
{\rm i} & 0 \\
0 & {\rm i}
\end{array}
\right)
&
L_1=  \frac{1}{2}\left(
\begin{array}{cc}
{\rm j} & 0 \\
0 & {\rm j}
\end{array}
\right)
&
L_2=  \frac{1}{2}\left(
\begin{array}{cc}
{\rm k} & 0 \\
0 & {\rm -k}
\end{array}
\right)
& 
L_3=  \frac{1}{2} \left(
\begin{array}{cc}
0 & {\rm k} \\
{\rm k} & 0
\end{array}
\right).
\end{array}
\end{equation}
Note that each of these matrices is anti-Hermitian (i.e. $J_i^{\dagger} = -J_i, K_i^{\dagger}=-K_i$, and $L_i^{\dagger}=-L_i$), as required by the  unitarity constraint (\ref{eq:lieconstraint}). Hence exponentiating any real linear combination of these matrices will yield a $2 \times 2$ unitary matrix over the split-quaternions. 

Now, invoking the defining relations (\ref{eq:sqalgebra}) for the split-quaternion algebra, we deduce the following commutation relations for $J_i$ and $K_i$:
\begin{equation}
\label{eq:LorentzLieAlgebra}
\begin{array}{llll}
\lbrack J_1,J_2 \rbrack= -J_3 & [K_1,K_2] = J_3 & [ J_1,K_2] = -K_3  &  [K_1,J_2]=-K_3 \\
\lbrack J_2,J_3 \rbrack= -J_1 & [K_2,K_3] = J_1 & [ J_2,K_3] = -K_1  &  [K_2,J_3]=-K_1 \\
\lbrack J_3,J_1 \rbrack= -J_2 & [K_3,K_1] = J_2 & [ J_3,K_1] = -K_2  &  [K_3,J_1]=-K_2
\end{array}
\end{equation}
All other commutation relations between $J_i$ and $K_i$ vanish. These generators therefore span a six dimensional subalgebra of the ten dimensional Lie algebra of U$(2,{\bf P})$.  In fact, we shall see in Section \ref{sec:u2pequiv} that this subalgebra corresponds to a representation of the Lorentz group. In particular, the $J_i$ correspond to rotations about the coordinate axes, while the $K_i$ correspond to Lorentz boosts.

As a side remark, note that the $L_i$ matrices satisfy the following equalities:
\begin{equation}
\label{eq:Lsquared}
\begin{array}{cccc}
(2L_0)^2 = {\bf -1} & (2L_1)^2 = {\bf +1}  & (2L_2)^2 = {\bf +1} & (2L_3)^2 = {\bf +1} .
\end{array}
\end{equation} 
Moreover, these four matrices also anti-commute with each other:
\begin{equation}
\label{eq:Lanticommute}
\left\{ L_i,L_j \right\}=L_i\cdot L_j + L_j \cdot L_i  = 0  \mbox{                 for $i\neq j$.}
\end{equation}
Properties (\ref{eq:Lsquared}) and (\ref{eq:Lanticommute}) above suggest that the four matrices $L_i$ are similar to the $\gamma^{\mu}$ matrices in the usual formulation of Dirac's equation. This analogy will be strengthened in Section \ref{sec:SQDiracEqn} where we write down a split-quaternionic version of the Dirac equation using the $L_i$ matrices as gamma matrices.

\subsection{The Isomorphism $u(2,{\bf P})$ $\cong$  $so(3,2;{\bf R})$} 
\label{sec:u2pequiv}

In this Section we show that the Lie algebra of the unitary group U$(2,{\bf P})$ is isomorphic to $so(3,2;{\bf R})$. To see why this is so, consider $5 \times 5$ matrices $M$ over the real numbers that satisfy the following constraint:
\begin{equation}
\label{eq:SO32defn}
M^T\cdot \eta \cdot M = \eta
\end{equation}
where $\eta$ is the diagonal matrix
\begin{equation}
\eta = \mbox{diag}(-1,+1,+1,+1,-1), 
\end{equation}
and
\begin{equation}
\label{eq:detcon}
\mbox{det}(M)=1.
\end{equation}
The set of all such matrices is isomorphic to the indefinite special orthogonal group $SO(3,2;{\bf R})$. Equivalently, we can think of such matrices $M$ as linear transformations acting on points $(T,X,Y,Z,U)$ in a five dimensional space that preserve the quadratic
\begin{equation}
-T^2+X^2+Y^2+Z^2-U^2.
\end{equation} 
In this notation, the subset of such matrices that act only on the four dimensional subspace of points $(T,X,Y,Z)$ while leaving the fifth coordinate $U$ unchanged forms a subgroup $SO(3,1;{\bf R})$, which is just the Lorentz group on four dimensional spacetime.
 
Now, consider elements $M$ $\in$  $SO(3,2;{\bf R})$ that may be written in the form
\begin{equation}
\label{eq:exp55matrix}
M = e^{Bs}
\end{equation}
where $B$ is a $5 \times 5$ matrix over the real numbers, and $s$ is an auxiliary real parameter. Substituting (\ref{eq:exp55matrix}) into the defining constraint (\ref{eq:SO32defn}), we obtain
\begin{equation}
\label{eq:SO32defn2}
e^{B^{T}s} \cdot \eta \cdot e^{Bs} = \eta.
\end{equation}
Differentiating both sides of equation (\ref{eq:SO32defn2}) with respect to the real variable $s$, and then setting this variable to zero, we obtain the following constraint on $B$:
\begin{equation}
\label{eq:so32algebra}
B^T \cdot \eta + \eta \cdot B = 0.
\end{equation}
It is now straightforward to show that any $5 \times 5$ matrix $B$ over the real numbers that satisfies the constraint  (\ref{eq:so32algebra}) may be written in the form
\begin{equation}
\label{eq:so32decomp}
B = \sum_{i=1}^{3}\alpha_i \tilde{J}_i + \sum_{i=1}^{3}\beta_i \tilde{K}_i + \sum_{i=0}^{3}\eta_i \tilde{L}_i
\end{equation} 
where the $\alpha_i,\beta_i$ and $\eta_i$ appearing above are real parameters, and the $\tilde{J}_i,\tilde{K}_i$ and $\tilde{L}_i$ are the real matrices specified below:

\begin{equation}
\label{eq:Jmatrices2}
\begin{array}{ccc}
\tilde{J}_1 = \begin{tiny} \left(
\begin{array}{ccccc}
0 & 0 & 0 & 0 & 0 \\
0 & 0 & 0 & 0 & 0 \\
0 & 0 & 0 & 1 & 0 \\
0 & 0 & -1 & 0 & 0 \\
0 & 0 & 0 & 0 & 0 \\
\end{array}
\right) 
\end{tiny}
&
\tilde{J}_2=  \begin{tiny} \left(
\begin{array}{ccccc}
0 & 0 & 0 & 0 & 0 \\
0 & 0 & 0 & -1 & 0 \\
0 & 0 & 0 & 0 & 0 \\
0 & 1 & 0 & 0 & 0 \\
0 & 0 & 0 & 0 & 0 \\
\end{array}
\right) 
\end{tiny}
&
\tilde{J}_3=  \begin{tiny} \left(
\begin{array}{ccccc}
0 & 0 & 0 & 0 & 0 \\
0 & 0 & 1 & 0 & 0 \\
0 & -1 & 0 & 0 & 0 \\
0 & 0 & 0 & 0 & 0 \\
0 & 0 & 0 & 0 & 0 \\
\end{array}
\right) 
\end{tiny}
\end{array} 
\end{equation}
\begin{equation}
\label{eq:Kmatrices2}
\begin{array}{ccc}
\tilde{K}_1 = \begin{tiny} \left(
\begin{array}{ccccc}
0 & 1 & 0 & 0 & 0 \\
1 & 0 & 0 & 0 & 0 \\
0 & 0 & 0 & 0 & 0 \\
0 & 0 & 0 & 0 & 0 \\
0 & 0 & 0 & 0 & 0 \\
\end{array}
\right)
\end{tiny}
&
\tilde{K}_2=  \begin{tiny} \left(
\begin{array}{ccccc}
0 & 0 & 1 & 0 & 0 \\
0 & 0 & 0 & 0 & 0 \\
1 & 0 & 0 & 0 & 0 \\
0 & 0 & 0 & 0 & 0 \\
0 & 0 & 0 & 0 & 0 \\
\end{array}
\right)
\end{tiny}
&
\tilde{K}_3=  \begin{tiny} \left(
\begin{array}{ccccc}
0 & 0 & 0 & 1 & 0 \\
0 & 0 & 0 & 0 & 0 \\
0 & 0 & 0 & 0 & 0 \\
1 & 0 & 0 & 0 & 0 \\
0 & 0 & 0 & 0 & 0 \\
\end{array}
\right)
\end{tiny}
\end{array}
\end{equation}
\begin{equation}
\label{eq:Lmatrices2}
\begin{array}{ccc}
\tilde{L}_0 = \begin{tiny} \left(
\begin{array}{ccccc}
0 & 0 & 0 & 0 & -1 \\
0 & 0 & 0 & 0 & 0 \\
0 & 0 & 0 & 0 & 0 \\
0 & 0 & 0 & 0 & 0 \\
1 & 0 & 0 & 0 & 0 \\
\end{array}
\right)
\end{tiny}
&
\tilde{L}_1=  \begin{tiny} \left(
\begin{array}{ccccc}
0 & 0 & 0 & 0 & 0 \\
0 & 0 & 0 & 0 & 1 \\
0 & 0 & 0 & 0 & 0 \\
0 & 0 & 0 & 0 & 0 \\
0 & 1 & 0 & 0 & 0 \\
\end{array}
\right)
\end{tiny}
&
\tilde{L}_2=  \begin{tiny} \left(
\begin{array}{ccccc}
0 & 0 & 0 & 0 & 0 \\
0 & 0 & 0 & 0 & 0 \\
0 & 0 & 0 & 0 & 1 \\
0 & 0 & 0 & 0 & 0 \\
0 & 0 & 1 & 0 & 0 \\
\end{array}
\right)
\end{tiny}
\end{array}
\end{equation}
\begin{equation}
\label{eq:Lmatrices3}
\begin{array}{c}
\tilde{L}_3 = \begin{tiny} \left(
\begin{array}{ccccc}
0 & 0 & 0 & 0 & 0 \\
0 & 0 & 0 & 0 & 0 \\
0 & 0 & 0 & 0 & 0 \\
0 & 0 & 0 & 0 & 1 \\
0 & 0 & 0 & 1 & 0 \\
\end{array} 
\right)
\end{tiny}
\end{array}
\end{equation}
Note that the ten generators $\tilde{J}_i, \tilde{K}_i$ and $\tilde{L_i}$ listed above are traceless, so exponentiating any real linear combination of these generators will yield a $5 \times 5$ real matrix satisfying the two defining conditions (\ref{eq:SO32defn}) and (\ref{eq:detcon}) of $SO(3,2;{\bf R})$.

Of course, the notation above is deliberately suggestive. The isomorphism between the two algebras $u(2,{\bf P})$ and $so(3,2;{\bf R})$ corresponds to making the following identifications:
\begin{equation}
\begin{array}{ccc}
J_i \leftrightarrow \tilde{J}_i  & K_i \leftrightarrow \tilde{K}_i  & L_i \leftrightarrow \tilde{L}_i ,
\end{array}
\end{equation}
where the $J_i,K_i$ and $L_i$ matrices were defined in Section \ref{sec:u2p}.
The explicit calculation of all the commutators to verify this equivalence is straightforward (though somewhat tedious), and omitted for the sake of brevity. 

Given this isomorphism of algebras, and viewing the $SO(3,2; {\bf R})$ group as acting on a five dimensional space of points $(T,X,Y,Z,U)$, we can see from inspection of (\ref{eq:Jmatrices2}) and (\ref{eq:Kmatrices2}) (and setting the auxiliary parameter to $s=1$) that the $\alpha_1, \alpha_2$ and $\alpha_3$ parameters appearing in the decompositions (\ref{eq:u2pdecomp}) and (\ref{eq:so32decomp}) correspond to rotation angles about the spatial $X,Y$ and $Z$ coordinate axes respectively.  The $\beta_1, \beta_2$ and $\beta_3$ parameters correspond to Lorentz boosts along the same $X,Y$ and $Z$ axes respectively. If any of the $\eta_i$ parameters are non-zero, then the fifth (time-like) coordinate $U$ is modified under an $SO(3,2; {\bf R})$ transformation. The $U$ coordinate is only left unchanged if we restrict transformations to the six dimensional Lorentz subgroup $SO(3,1; {\bf R})$.

\section{The Lorentz Transformation on Spinors}
\label{sec:SQSpinors}
\subsection{Dirac 4-Spinors}
\label{sec:diracdefn}
First some definitions. We write $\vec{\sigma}  =(\sigma^1,\sigma^2,\sigma^3)$ where 
\begin{equation}
\begin{array}{ccc}
\sigma^1 = \left(
\begin{array}{cc}
0 & 1 \\
1 & 0
\end{array}
\right) 
&
\sigma^2 = \left(
\begin{array}{cc}
0 & {\rm -i} \\
{\rm i} & 0
\end{array}
\right) 
&
\sigma^3 = \left(
\begin{array}{cc}
1 & 0 \\
0 & -1
\end{array}
\right) 
\end{array}
\end{equation}
are the well known Pauli spin matrices. Now if $\psi_{W}$ is a 4-spinor in the chiral (i.e. Weyl) representation, then a Lorentz transformation takes the form
\begin{equation}
\label{eq:lorentztransform}
\psi_{W} \rightarrow 
\begin{large}
\left(
\begin{array}{cc}
e^{\frac{\rm{ i}}{2}\vec{\alpha} \cdot \vec{\sigma}+\frac{1}{2}\vec{\beta} \cdot \vec{\sigma}}  & 0 \\
0 & e^{\frac{\rm{i}}{2}\vec{\alpha} \cdot \vec{\sigma}- \frac{1}{2}\vec{\beta} \cdot \vec{\sigma}}
\end{array}
\right)
\end{large}
\cdot \psi_{W}
\end{equation}
where $\vec{\alpha}=(\alpha_1,\alpha_2,\alpha_3)$ represents the three rotation angles about the coordinate axes, while $\vec{\beta}=(\beta_1,\beta_2,\beta_3)$ denotes the three Lorentz boosts along the same coordinate axes. A proof of this can be found in any standard text book on relativistic quantum field theory.

Note that expression (\ref{eq:lorentztransform}) can be written in the form 
\begin{equation}
\label{eq:LTJKform}
\psi_{W} \rightarrow \mbox{exp} \left( \sum_{i=1}^3 \alpha_i \hat{J_i}+\sum_{i=1}^3 \beta_i \hat{K_i} \right) \cdot \psi_{W}
\end{equation}
where the six $4 \times 4$ complex matrices $\hat{J_i}$ and $\hat{K_i}$ appearing above are defined by writing
\begin{equation}
\label{eq:Jhatmatrices}
\begin{array}{ccc}
\hat{J_1} = \frac{1}{2} \left(
\begin{array}{cc}
{\rm i} \sigma^1 & 0 \\
0 & {\rm i}\sigma^1
\end{array}
\right)
&
\hat{J_2} = \frac{1}{2}\left(
\begin{array}{cc}
{\rm i}\sigma^2 & 0 \\
0 & {\rm i}\sigma^2
\end{array}
\right)
&
\hat{J_3} = \frac{1}{2}\left(
\begin{array}{cc}
{\rm i}\sigma^3 & 0 \\
0 & {\rm i}\sigma^3
\end{array}
\right)
\end{array}
\end{equation}
\begin{equation}
\label{eq:Khatmatrices}
\begin{array}{ccc}
\hat{K_1} = \frac{\rm 1}{2}\left(
\begin{array}{cc}
\sigma^1 & 0 \\
0 & -\sigma^1
\end{array}
\right)
&
\hat{K_2} = \frac{\rm 1}{2}\left(
\begin{array}{cc}
\sigma^2 & 0 \\
0 & -\sigma^2
\end{array}
\right)
&
\hat{K_3} = \frac{\rm 1}{2}\left(
\begin{array}{cc}
\sigma^3 & 0 \\
0 & -\sigma^3
\end{array}
\right)
\end{array}
\end{equation}
It is now straightforward to show that the $\hat{J_i}$ and $\hat{K_i}$ matrices satisfy the algebra of commutation relations (\ref{eq:LorentzLieAlgebra}) after making the identification $J_i \leftrightarrow \hat{J_i}$ and $K_i \leftrightarrow \hat{K_i}$. 

The existence of an isomorphism between the Lie algebra of the Lorentz group generated by the $\hat{J_i}$ and $\hat{K_i}$ matrices, and the Lie algebra generated by the $2 \times 2$ split-quaternionic matrices $J_i$ and $K_i$, suggests that a four component Dirac spinor might be equivalent to a two component split-quaternionic spinor, since the number of real degrees of freedom match (in this case eight).

To establish this equivalence, we will first need to study the  transformation properties for each of the eight real components of a Dirac 4-spinor. Let's first assume that the $\alpha_i$ and $\beta_i$ parameters appearing  in the Lorentz transformation  are infinitesimally small so that we can write (\ref{eq:LTJKform}) as follows:
\begin{equation}
\label{eq:LTJKformII}
\psi_W \rightarrow  \left({\bf 1}+ \sum_{i=1}^3 \alpha_i \hat{J_i}+\sum_{i=1}^3 \beta_i \hat{K_i} \right) \cdot \psi_W
\end{equation}
If we also define  
\begin{equation}
\psi_W = \left(
\begin{array}{c}
x_1 + {\rm i} y_1 \\
x_2 + {\rm i} y_2 \\
x_3 + {\rm i} y_3 \\
x_4 + {\rm i} y_4  
\end{array}
\right)
\end{equation}
where the $x_i$ and $y_i$ are the eight real components of the 4-spinor $\psi_W$, then the Lorentz transformation (\ref{eq:LTJKformII}) is equivalent to the the following transformations of the eight real components $x_i$ and $y_i$:
\begin{equation}
\label{eq:xtransform}
\begin{array}{c}
x_1 \rightarrow x_1 +\frac{1}{2}(-\alpha_1 y_2+\alpha_2 x_2-\alpha_3 y_1  +\beta_1 x_2 + \beta_2 y_2 + \beta_3 x_1) \\
x_2 \rightarrow x_2 +\frac{1}{2}(-\alpha_1 y_1-\alpha_2 x_1+\alpha_3 y_2  +\beta_1 x_1 - \beta_2 y_1 - \beta_3 x_2) 
\\
x_3 \rightarrow x_3 +\frac{1}{2}(-\alpha_1 y_4+\alpha_2 x_4-\alpha_3 y_3  -\beta_1 x_4 - \beta_2 y_4 - \beta_3 x_3) 
\\
x_4 \rightarrow x_4 +\frac{1}{2}(-\alpha_1 y_3-\alpha_2 x_3+\alpha_3 y_4  -\beta_1 x_3 + \beta_2 y_3 + \beta_3 x_4) 
\\
\end{array}
\end{equation} 

\begin{equation}
\label{eq:ytransform}
\begin{array}{c}
y_1 \rightarrow y_1 +\frac{1}{2}(\alpha_1 x_2+\alpha_2 y_2+\alpha_3 x_1  +\beta_1 y_2 - \beta_2 x_2 + \beta_3 y_1) \\
y_2 \rightarrow y_2 +\frac{1}{2}(\alpha_1 x_1-\alpha_2 y_1-\alpha_3 x_2  +\beta_1 y_1 + \beta_2 x_1 - \beta_3 y_2) 
\\
y_3 \rightarrow y_3 +\frac{1}{2}(\alpha_1 x_4+\alpha_2 y_4+\alpha_3 x_3  -\beta_1 y_4 + \beta_2 x_4 - \beta_3 y_3) 
\\
y_4 \rightarrow y_4 +\frac{1}{2}(\alpha_1 x_3-\alpha_2 y_3-\alpha_3 x_4  -\beta_1 y_3 -\beta_2 x_3 + \beta_3 y_4) 
\\
\end{array}
\end{equation} 
Since (\ref{eq:xtransform}) and (\ref{eq:ytransform}) represent the transformation properties  of a 4-spinor in the chiral (i.e. Weyl) representation, we will need a few extra steps to deduce the corresponding relations for a 4-spinor in the standard (i.e. Dirac) representation. 

 First note that if $\psi_{W}$ is a 4-spinor in the Weyl representation, while $\psi$ is a 4-spinor in the Dirac representation, then we can connect the two spinors by the following invertible transformation:
\begin{equation}
\label{eq:weyltodirac}
\psi = \frac{1}{\sqrt{2}}
\left(
\begin{array}{cc}
{\bf 1} & {\bf 1} \\
{\bf 1} & -{\bf 1}
\end{array}
\right) \cdot \psi_W
\end{equation}
where $\bf{1}$ corresponds to the $2 \times 2$ identity matrix. If we define $\psi_W$ and $\psi$ in terms of their eight real components by writing
\begin{equation}
\label{eq:spinordefns}
\begin{array}{ccc}
\psi _W= \left(
\begin{array}{c}
x_1 + {\rm i} y_1 \\
x_2 + {\rm i} y_2 \\
x_3 + {\rm i} y_3 \\
x_4 + {\rm i} y_4  
\end{array}
\right) &
\mbox{ and } &
\psi = \left(
\begin{array}{c}
u_1 + {\rm i} v_1 \\
u_2 + {\rm i} v_2 \\
u_3 + {\rm i} v_3 \\
u_4 + {\rm i} v_4  
\end{array}
\right) ,
\end{array}
\end{equation}
then (\ref{eq:weyltodirac}) allows us to write the eight real components $u_i$ and $v_i$ of $\psi$ in terms of the components $x_i$ and $y_i$ of $\psi_W$:
\begin{equation}
\label{eq:uvxy}
\begin{array}{cc}
u_1 = \frac{1}{\sqrt{2}}(x_1+x_3) & v_1 = \frac{1}{\sqrt{2}}(y_1+y_3) \\
u_2 = \frac{1}{\sqrt{2}}(x_2+x_4) & v_2 = \frac{1}{\sqrt{2}}(y_2+y_4) \\
u_3 = \frac{1}{\sqrt{2}}(x_1-x_3) & v_3 = \frac{1}{\sqrt{2}}(y_1-y_3) \\
u_4 = \frac{1}{\sqrt{2}}(x_2-x_4) & v_4 = \frac{1}{\sqrt{2}}(y_2-y_4)
\end{array}
\end{equation}
Now, by combining identities (\ref{eq:uvxy}) with relations (\ref{eq:xtransform}) and (\ref{eq:ytransform}), we deduce that under a Lorentz transformation, the eight real components $u_i$ and $v_i$ of a 4-spinor $\psi$ in the Dirac representation transform as follows:

\begin{equation}
\label{eq:utransform}
\begin{array}{c}
u_1 \rightarrow u_1 +\frac{1}{2}(-\alpha_1 v_2+\alpha_2 u_2-\alpha_3 v_1  +\beta_1 u_4 + \beta_2 v_4 + \beta_3 u_3) \\
u_2 \rightarrow u_2 +\frac{1}{2}(-\alpha_1 v_1-\alpha_2 u_1+\alpha_3 v_2  +\beta_1 u_3 - \beta_2 v_3 - \beta_3 u_4) 
\\
u_3 \rightarrow u_3 +\frac{1}{2}(-\alpha_1 v_4+\alpha_2 u_4-\alpha_3 v_3  +\beta_1 u_2 + \beta_2 v_2 + \beta_3 u_1) 
\\
u_4 \rightarrow u_4 +\frac{1}{2}(-\alpha_1 v_3-\alpha_2 u_3+\alpha_3 v_4  +\beta_1 u_1 - \beta_2 v_1 - \beta_3 u_2) 
\\
\end{array}
\end{equation} 
\begin{equation}
\label{eq:vtransform}
\begin{array}{c}
v_1 \rightarrow v_1 +\frac{1}{2}(\alpha_1 u_2+\alpha_2 v_2+\alpha_3 u_1  +\beta_1 v_4 - \beta_2 u_4 + \beta_3 v_3) 
\\
v_2 \rightarrow v_2 +\frac{1}{2}(\alpha_1 u_1-\alpha_2 v_1-\alpha_3 u_2  +\beta_1 v_3 + \beta_2 u_3 - \beta_3 v_4) 
\\
v_3 \rightarrow v_3 +\frac{1}{2}(\alpha_1 u_4+\alpha_2 v_4+\alpha_3 u_3  +\beta_1 v_2 - \beta_2 u_2 + \beta_3 v_1) 
\\
v_4 \rightarrow v_4 +\frac{1}{2}(\alpha_1 u_3-\alpha_2 v_3-\alpha_3 u_4  +\beta_1 v_1 + \beta_2 u_1 - \beta_3 v_2) 
\\
\end{array}
\end{equation} 
As before, we have assumed that the $\alpha_i$ and $\beta_i$ parameters are infinitesimally small, so all higher order corrections are ignored. 

The relations (\ref{eq:utransform}) and (\ref{eq:vtransform}) will be convenient when proving the equivalence between Dirac 4-spinors and split-quaternionic 2-spinors in Section \ref{sec:equivspinors}. Towards this end, we discuss the transformation properties of split-quaternionic 2-spinors next.

\subsection{Split-Quaternionic 2-Spinors}
\label{sec:defnsqspinor}
Let us define a 2-spinor $\Psi$ over the split-quaternions as follows:
\begin{equation}
\label{eq:SQ2spinor}
\Psi = 
\left(
\begin{array}{c}
a_1 + {\rm i} a_2 + {\rm j} a_3 + {\rm k} a_4 \\
b_1 + {\rm i} b_2 + {\rm j} b_3 + {\rm k} b_4  
\end{array}
\right)
\end{equation}
where the $a_i$ and $b_i$ are the eight real components of $\Psi$. Thus $\Psi$ is simply an ordered pair of split-quaternions. 

Now in Section \ref{sec:u2p} we showed there exist $2 \times 2$ split-quaternionic matrices $J_i$ and $K_i$ that satisfy the Lorentz Lie algebra (\ref{eq:LorentzLieAlgebra}). Consequently, under a Lorentz transformation, the split-quaternionic 2-spinor defined by (\ref{eq:SQ2spinor}) will transform as follows:
\begin{equation}
\label{eq:LTJKSQform}
\Psi \rightarrow \mbox{exp} \left( \sum_{i=1}^3 \alpha_i J_i+\sum_{i=1}^3 \beta_i K_i \right) \cdot \Psi
\end{equation}
where the real parameters $\alpha_i$ and $\beta_i$ are the rotation angles and Lorentz boost parameters respectively. As before, if we assume these real parameters are infinitesimally small, then (\ref{eq:LTJKSQform}) takes the simpler form
\begin{equation}
\label{eq:LTJKSQform2}
\Psi \rightarrow  \left({\bf 1} + \sum_{i=1}^3 \alpha_i J_i+\sum_{i=1}^3 \beta_i K_i \right) \cdot \Psi
\end{equation}
In terms of the eight real components $a_i$ and $b_i$ of $\Psi$ appearing in definition (\ref{eq:SQ2spinor}), the infinitesimal Lorentz transformation (\ref{eq:LTJKSQform2}) is equivalent to the following explicit transformations:
\begin{equation}
\label{eq:atransform}
\begin{array}{c}
a_1 \rightarrow a_1 +\frac{1}{2}(\alpha_1 b_1-\alpha_2 b_2+\alpha_3 a_2  -\beta_1 a_4 + \beta_2 a_3 + \beta_3 b_3)
\\
a_2 \rightarrow a_2 +\frac{1}{2}(\alpha_1 b_2+\alpha_2 b_1-\alpha_3 a_1  -\beta_1 a_3 - \beta_2 a_4 - \beta_3 b_4) 
\\
a_3 \rightarrow a_3 +\frac{1}{2}(\alpha_1 b_3-\alpha_2 b_4+\alpha_3 a_4  -\beta_1 a_2 + \beta_2 a_1 + \beta_3 b_1) 
\\
a_4 \rightarrow a_4 +\frac{1}{2}(\alpha_1 b_4+\alpha_2 b_3-\alpha_3 a_3  -\beta_1 a_1 - \beta_2 a_2 - \beta_3 b_2) 
\\
\end{array}
\end{equation} 
\begin{equation}
\label{eq:btransform}
\begin{array}{c}
b_1 \rightarrow b_1 +\frac{1}{2}(-\alpha_1 a_1-\alpha_2 a_2-\alpha_3 b_2  -\beta_1 b_4 - \beta_2 b_3 + \beta_3 a_3)
\\
b_2 \rightarrow b_2 +\frac{1}{2}(-\alpha_1 a_2+\alpha_2 a_1+\alpha_3 b_1  -\beta_1 b_3 + \beta_2 b_4 - \beta_3 a_4) 
\\
b_3 \rightarrow b_3 +\frac{1}{2}(-\alpha_1 a_3-\alpha_2 a_4-\alpha_3 b_4  -\beta_1 b_2 - \beta_2 b_1 + \beta_3 a_1) 
\\
b_4 \rightarrow b_4 +\frac{1}{2}(-\alpha_1 a_4+\alpha_2 a_3+\alpha_3 b_3  -\beta_1 b_1 + \beta_2 b_2 - \beta_3 a_2) 
\\
\end{array}
\end{equation} 
Recall that a Dirac 4-spinor has eight real components, and in the Dirac representation, these eight real components transform in accordance with relations (\ref{eq:utransform}) and (\ref{eq:vtransform}). 

A split-quaternionic 2-spinor also has eight real components, and under the same infinitesimal Lorentz transformation, these eight real components transform in accordance with relations  (\ref{eq:atransform}) and (\ref{eq:btransform}). 

In the next Section, we show that under a suitable identification of components, these two sets of relations are equivalent, and so split-quaternionic 2-spinors are an equivalent way of representing Dirac 4-spinors.

\subsection{An Equivalence}
\label{sec:equivspinors}
Let $\psi$ be a 4-spinor in the Dirac representation  as defined in (\ref{eq:spinordefns}), with eight real components $u_i$ and $v_i$ transforming in accordance with relations  (\ref{eq:utransform}) and (\ref{eq:vtransform}) under an infinitesimal Lorentz transformation. 

In addition, let $\Psi$ be a split-quaternionic 2-spinor as defined in (\ref{eq:SQ2spinor}) with eight real components $a_i$ and $b_i$ transforming in accordance with relations  (\ref{eq:atransform}) and (\ref{eq:btransform}) under the same infinitesimal Lorentz transformation.

Then the transformations (\ref{eq:utransform}) and (\ref{eq:vtransform}) governing the $u_i$ and $v_i$ components are equivalent to relations (\ref{eq:atransform}) and (\ref{eq:btransform}) governing the $a_i$ and $b_i$ components if we make the following identification between components:
\begin{equation}
\label{eq:equivofreps}
\begin{array}{ll}
u_1 \leftrightarrow +a_1 & v_1 \leftrightarrow -a_2 \\
u_2 \leftrightarrow -b_2 & v_2 \leftrightarrow -b_1 \\
u_3 \leftrightarrow +b_3 & v_3 \leftrightarrow +b_4 \\
u_4 \leftrightarrow -a_4 & v_4 \leftrightarrow +a_3
\end{array}
\end{equation}
In spinor notation, the identifications specified by (\ref{eq:equivofreps}) above takes the following form:
\begin{equation}
\label{eq:spinorequivII}
\begin{array}{ccc}
\Psi = 
\left(
\begin{array}{c}
a_1 + {\rm i} a_2 + {\rm j} a_3 + {\rm k} a_4 \\
b_1 + {\rm i} b_2 + {\rm j} b_3 + {\rm k} b_4  
\end{array}
\right) &
\leftrightarrow &
\psi = \left(
\begin{array}{c}
 \mbox{\space \space} a_1- {\rm i} a_2 \\
-b_2- {\rm i} b_1 \\
\mbox{\space \space} b_3 + {\rm i} b_4 \\
-a_4 + {\rm i} a_3  
\end{array}
\right)
\end{array}
\end{equation}
At this point, we remark that the mapping (\ref{eq:equivofreps}) between components is not unique. There are known to be a total of eight such mappings. This was first hinted at in the earlier work on hyperbolic spinors \cite{AntonuccioSpinorPaperI}, but in the present case, we are only interested in showing that such a mapping exists. The existence of additional mappings, and relations between them, is outside the scope of this  paper.

Now, recall that the $\gamma^0$ matrix in the Dirac representation is given by
\begin{equation}
\label{eq:gamma0}
\gamma^0 = 
\left(
\begin{array}{cc}
{\bf 1} & {\bf 0} \\
{\bf 0} & -{\bf 1}
\end{array}
\right) 
\end{equation}
where {\bf 1} is the $2 \times 2$ identity matrix. Thus, by virtue of the identification (\ref{eq:spinorequivII}) between spinor formalisms, we deduce that
\begin{eqnarray}
\label{eq:psibarcalc}
\overline{\psi} \psi & = & \psi^{\dagger} \gamma^0 \psi \nonumber \\
                                                 & = & (a_1)^2+(a_2)^2-(a_3)^2-(a_4)^2 + (b_1)^2+(b_2)^2-(b_3)^2-(b_4)^2 \nonumber \\
                                                 & = & \Psi^{\dagger} \cdot \Psi
\end{eqnarray}
Thus, in the split-quaternionic formalism, there is no need for a $\gamma^0$ matrix in order to construct Lorentz invariants, since the Lorentz group is unitary with respect to the split-quaternion algebra. 

As a concluding remark, note that $\Psi^{\dagger} \cdot \Psi$ is naturally invariant under the unitary group U$(2,{\bf P})$, which we now know to be locally isomorphic to $SO(3,2;{\bf R})$. The calculation (\ref{eq:psibarcalc}) makes it clear that the Lorentz invariant quantity $\overline{\psi}\psi$ is in fact invariant under $SO(3,2;{\bf R})$ transformations which contains the Lorentz group as a subgroup.

\section{The Dirac Equation}
\label{sec:SQDiracEqn}
\subsection{Dirac 4-Spinor Equation}
We wish to strengthen the identification between Dirac 4-spinors $\psi$ and split-quaternionic 2-spinors $\Psi$ by showing there exists an equation governing the time evolution of $\Psi$ that reduces to the usual Dirac equation after making the identification (\ref{eq:spinorequivII}) between spinor formalisms.

To accomplish this, we first need to revisit the Dirac equation in component form. Let $\gamma_W^{\mu}$ denote the four gamma matrices for the Dirac equation in the Weyl representation. Explicitly, we may write
\begin{equation}
\label{eq:gammaWeyl}
\begin{array}{ccc}
\gamma_W^0 = 
\left(
\begin{array}{cc}
{\bf 0} & {\bf 1} \\
{\bf 1} & {\bf 0}
\end{array}
\right)
&
\mbox{ and}
&
\gamma_W^i = 
\left(
\begin{array}{cc}
{\bf 0} & -\sigma^i \\
\sigma^i & {\bf 0}
\end{array}
\right)
\end{array}
\end{equation}
where $i=1,2,3$. In the above, ${\bf 1}$ denotes the $2 \times 2$ identity matrix, while the Pauli matrices $\sigma^i$ were defined earlier in Section \ref{sec:diracdefn}.

If $\psi_W$ is a 4-spinor in the Weyl representation, then the corresponding Dirac equation governing its time evolution is given by
\begin{equation}
\left( {\rm i}\gamma_W^{\mu} \partial_{\mu} - m \right) \psi_W = 0.
\end{equation}
Now, let $\gamma^{\mu}$ denote the four gamma matrices in the Dirac representation. There is a simple way to connect these gamma matrices to those in the Weyl representation; namely
\begin{equation}
\label{eq:diracgamma}
\gamma^{\mu} = U \gamma_W^{\mu} U^{-1} \mbox{\space \space for $\mu = 0,1,2,3$,}
\end{equation}
where $U$ is a $4 \times 4$ unitary matrix defined by
\begin{equation}
\label{eq:unitaryconnection}
U = \frac{1}{\sqrt{2}}
\left(
\begin{array}{cc}
{\bf 1} & {\bf 1} \\
{\bf 1} & -{\bf 1}
\end{array}
\right) .
\end{equation}
Combining definitions (\ref{eq:gammaWeyl}),(\ref{eq:diracgamma}) and (\ref{eq:unitaryconnection}), we deduce the following explicit expressions for the gamma matrices $\gamma^{\mu}$ in the Dirac representation:
\begin{equation}
\begin{array}{ccc}
\gamma^0 = 
\left(
\begin{array}{cc}
{\bf 1} & {\bf 0} \\
{\bf 0} & -{\bf 1}
\end{array}
\right)
&
\mbox{ and}
&
\gamma^i = 
\left(
\begin{array}{cc}
{\bf 0} & \sigma^i \\
-\sigma^i & {\bf 0}
\end{array}
\right).
\end{array}
\end{equation}
So if $\psi$ is a 4-spinor in the Dirac representation, it's equation of motion is governed by 
\begin{equation}
\label{eq:standardDEqn}
\left( {\rm i}\gamma^{\mu} \partial_{\mu} - m \right) \psi = 0.
\end{equation}
Let us now make the explicit definition 
\begin{equation}
\label{eq:psidefnuv}
\psi = \left(
\begin{array}{c}
u_1 + {\rm i} v_1 \\
u_2 + {\rm i} v_2 \\
u_3 + {\rm i} v_3 \\
u_4 + {\rm i} v_4  
\end{array}
\right)
\end{equation}
where the $u_i$ and $v_i$ are the eight real components of $\psi$. Substituting (\ref{eq:psidefnuv})  into the Dirac equation (\ref{eq:standardDEqn}), we deduce the following equations governing the eight real components $u_i$ and $v_i$: 
\begin{equation}
\label{eq:ueqns}
\begin{array}{c}
-\partial_0v_1-\partial_1 v_4 +\partial_2 u_4 - \partial_3 v_3 = m u_1 \\
-\partial_0v_2-\partial_1 v_3 -\partial_2 u_3 + \partial_3 v_4 = m u_2 \\
+\partial_0v_3+\partial_1 v_2 -\partial_2 u_2 + \partial_3 v_1 = m u_3 \\
+\partial_0v_4+\partial_1 v_1 +\partial_2 u_1 - \partial_3 v_2 = m u_4 
\end{array}
\end{equation}
\begin{equation}
\label{eq:veqns}
\begin{array}{c}
+\partial_0 u_1+\partial_1 u_4 +\partial_2 v_4 + \partial_3 u_3 = m v_1 \\
+\partial_0 u_2+\partial_1 u_3 -\partial_2 v_3 - \partial_3 u_4 = m v_2 \\
-\partial_0 u_3-\partial_1 u_2 -\partial_2 v_2 - \partial_3 u_1 = m v_3 \\
-\partial_0u_4-\partial_1 u_1 +\partial_2 v_1 + \partial_3 u_2 = m v_4 
\end{array}
\end{equation}
The spinor component equations (\ref{eq:ueqns}) and (\ref{eq:veqns}) arising from the Dirac equation will turn out to be equivalent to a set of equations arising from a very natural split-quaternionic 2-spinor equation. We discuss this topic next.

\subsection{Split-Quaternionic 2-Spinor Equation}
In order to construct a split-quaternionic version of the Dirac equation, recall that the four $2 \times 2$ split-quaternionic matrices $L_i$ appearing in (\ref{eq:Lmatrices}) satisfied an algebra that was analogous to the Dirac algebra of the four $\gamma^{\mu}$ matrices, as evidenced by the relations (\ref{eq:Lsquared}) and (\ref{eq:Lanticommute}). 

In order to make this analogy more precise, we introduce four new $2 \times 2$ split-quaternionic matrices $H^{\mu}$ by defining
\begin{equation}
\label{eq:HMatricesII}
\begin{array}{cccc}
H^0 = \left(
\begin{array}{cc}
{\rm i} & 0 \\
0 & {\rm i}
\end{array}
\right) &
H^1 = \left(
\begin{array}{cc}
{\rm j} & 0 \\
0 & {\rm j}
\end{array}
\right) &
H^2 = \left(
\begin{array}{cc}
{\rm k} & 0 \\
0 & {\rm -k}
\end{array}
\right) &
H^3 = \left(
\begin{array}{cc}
0 & {\rm k} \\
{\rm k} & 0
\end{array}
\right) .
\end{array}
\end{equation}
By virtue of identities (\ref{eq:Lsquared}) and (\ref{eq:Lanticommute}), or by direct substitution, it follows that the $H^{\mu}$ matrices satisfy the following algebra of anti-commutation relations:
\begin{equation}
\{  H^{\mu},H^{\nu} \}=H^{\mu}\cdot H^{\nu}+H^{\nu}\cdot H^{\mu}= 2g^{\mu \nu} {\bf 1}
\end{equation}
where ${\bf 1}$ is the $2 \times 2$ identity matrix, and $(g^{\mu \nu})$ is the diagonal matrix defined by 
\begin{equation}
 (g^{\mu \nu}) = \mbox{diag}(-1,+1,+1,+1).
\end{equation}

We are now ready to define our promised equation. Let $\Psi$ be a 2-spinor over the split-quaternions as defined in Section \ref{sec:defnsqspinor}. Then the equation of motion governing $\Psi$ is given by
\begin{equation}
\label{eq:THEeqn}
\left(H^{\mu} \partial_{\mu} + m \right) \cdot \Psi = 0.
\end{equation}
If we write the explicit definition 
\begin{equation}
\label{eq:SQ2spinorII}
\Psi = 
\left(
\begin{array}{c}
a_1 + {\rm i} a_2 + {\rm j} a_3 + {\rm k} a_4 \\
b_1 + {\rm i} b_2 + {\rm j} b_3 + {\rm k} b_4  
\end{array}
\right)
\end{equation}
where the $a_i$ and $b_i$ are the eight real components of $\Psi$, then equation (\ref{eq:THEeqn}) is equivalent to the following set of equations:
\begin{equation}
\label{eq:aeqns}
\begin{array}{c}
+\partial_0a_2-\partial_1 a_3 -\partial_2 a_4 - \partial_3 b_4 = m a_1 \\
-\partial_0a_1+\partial_1 a_4 -\partial_2 a_3 - \partial_3 b_3 = m a_2 \\
+\partial_0a_4-\partial_1 a_1 -\partial_2 a_2 - \partial_3 b_2 = m a_3 \\
-\partial_0a_3+\partial_1 a_2 -\partial_2 a_1 - \partial_3 b_1 = m a_4 
\end{array}
\end{equation}
\begin{equation}
\label{eq:beqns}
\begin{array}{c}
+\partial_0b_2-\partial_1 b_3 +\partial_2 b_4 - \partial_3 a_4 = m b_1 \\
-\partial_0b_1+\partial_1 b_4 +\partial_2 b_3 - \partial_3 a_3 = m b_2 \\
+\partial_0b_4-\partial_1 b_1 +\partial_2 b_2 - \partial_3 a_2 = m b_3 \\
-\partial_0b_3+\partial_1 b_2 +\partial_2 b_1 - \partial_3 a_1 = m b_4 
\end{array}
\end{equation}
In the next Section, we show that the equations of motion (\ref{eq:aeqns}) and (\ref{eq:beqns}) are just the usual Dirac equation in disguise.

\subsection{An Equivalence}

We are now in a position to state an equivalence between the equations of motion (\ref{eq:ueqns}) and (\ref{eq:veqns}) governing the eight real components $u_i$ and $v_i$ of a 4-spinor $\psi$ in the Dirac representation, and the equations of motion (\ref{eq:aeqns}) and (\ref{eq:beqns}) governing the eight real components $a_i$ and $b_i$ of a split-quaternionic 2-spinor $\Psi$. Namely, under the identification of components specified by relations (\ref{eq:equivofreps}) (or equivalently, under the identification of spinors stated in (\ref{eq:spinorequivII})), these two sets of equations are identical. 

Hence the Dirac equation 
\begin{equation}
\label{eq:standardDEqnII}
\left( {\rm i}\gamma^{\mu} \partial_{\mu} - m \right) \psi = 0
\end{equation}
governing the 4-spinor $\psi$ in the Dirac representation has an equivalent representation in terms of the split-quaternionic 2-spinor equation
 \begin{equation}
\label{eq:THEeqnII}
\left(H^{\mu} \partial_{\mu} + m \right) \cdot \Psi = 0.
\end{equation}
The split-quaternionic representation (\ref{eq:THEeqnII}) above is particularly simple, since three of the $H^{\mu}$ matrices are diagonal. The non-diagonal matrix $H^3$ provides the coupling between the two split-quaternionic components of the 2-spinor $\Psi$. For the special case
\begin{equation}
\partial_3 \Psi = 0,
\end{equation}
these two components of $\Psi$ completely decouple.

\section{Concluding Remarks}
\label{sec:Conclusions}
In this paper, we gave an explicit example of a finite unitary representation of the Lorentz group in terms of $2 \times 2$ matrices over the split-quaternions. We also showed that this representation of the Lorentz group is naturally embedded inside the ten dimensional unitary group U$(2,{\bf P})$, which is locally isomorphic to $SO(3,2;{\bf R})$. In this new formalism, the $SO(3,2;{\bf R})$ symmetry of the Lorentz invariant scalar $\overline{\psi} \psi$ is particularly transparent.

We also showed that the standard action of the Lorentz group on a Dirac 4-spinor is entirely equivalent to a split-quaternionic representation of the Lorentz group acting on a split-quaternionic 2-spinor. Moreover, the Dirac equation governing the 4-spinor has an equivalent formulation in terms of this same split-quaternionic 2-spinor.

Expressing a Dirac 4-spinor as a 2-spinor object has been attempted before by Hucks \cite{Hucks}, but the representation here differs in a crucial respect. The algebra of the split-quaternions is non-commutative, and gives rise to a natural unitary representation of the Lorentz group. In contrast, the hyperbolic complex algebra investigated by Hucks is a commutative extension of the complex numbers. 

It should be pointed out that a study of the spinor representation of the Lorentz group in terms of $2 \times 2$ matrices over Hamilton's quaternions (and corresponding Dirac theory) is summarised in Morita's paper \cite{Morita}. In this work, Morita introduces an additional commuting $\sqrt{-1}$ factor along with the usual elements of the quaternion algebra, which amounts to `complexifying' his quaternionic variables when writing down Dirac's equation. In contrast, the properties of the split-quaternion algebra allow one to write down a real representation of the Dirac equation over the split-quaternions.

In the context of the Standard Model, Morita established an interesting connection between quaternions, non-commutative geometry and the electro-weak bosonic sector of Weinberg-Salam theory \cite{MoritaII}. This suggests that a possible direction of research is to parallel Morita's analysis in terms of split-quaternionic variables. 

Note that the Dirac equation coupled to a background electromagnetic field has been shown to assume a particularly simple form in the context of split-octonions \cite{Merab}, which contains the split-quaternions as a special case. In this article, we have avoided introducing such non-associative algebras to describe the Dirac equation, but we do not rule out that a more profound understanding may require the use of such algebraic systems. In fact, in the context of Geometric Algebra, a deeper formulation of the Dirac equation without any reference to matrices has already been considered \cite{Hestenes},\cite{Rodrigues},\cite{RodriguesII}. 

It is worth remarking that considerable effort has been made over the years to understand the equations of physics -- and spacetime itself -- by treating spinors as the fundamental degrees of freedom of a physical system. This idea was  pioneered by Penrose in his twistor theory program \cite{Penrose}. From this point of view, the complex numbers and complex analysis play a prominent role. In light of the equivalence established in this paper, it is tempting to generalise Penrose's program by investigating further the role of split-quaternions and split-quaternionic analysis in modern physics.

\end{document}